\documentclass[lettersize,journal]{IEEEtran}
\usepackage{cite}
\usepackage{amsmath,amssymb,amsfonts}
\usepackage{graphicx}
\usepackage{array}
\usepackage[caption=false,font=normalsize,labelfont=sf,textfont=sf]{subfig}
\usepackage{textcomp}
\usepackage{bbm}
\usepackage[]{algorithmicx}
\usepackage{algpseudocode,algorithm}
\usepackage{mathtools}
\usepackage{xcolor}
\def\R{{\mathbb{R}}}

\def\O{{\mathcal{O}}}

\newcommand{\norm}[1]{\left\|{#1}\right\|}
\def\BibTeX{{\rm B\kern-.05em{\sc i\kern-.025em b}\kern-.08em
    T\kern-.1667em\lower.7ex\hbox{E}\kern-.125emX}}

\begin{document}
\title{Nonlinear Waveform Inversion for Quantitative Ultrasound}

\author{Avner Shultzman, \IEEEmembership{Student Member, IEEE} and Yonina C. Eldar, \IEEEmembership{Fellow, IEEE}
\thanks{This project has received funding from the European Research Council (ERC) under the European Union's Horizon 2020 research and innovation programme (grant agreement No. 101000967) and from the Manya Igel Centre for Biomedical Engineering and Signal Processing.}}

\markboth{IEEE Transactions on Computational Imaging, VOL. XX, NO. XX, XXXX 2022}
{SHULTZMAN and ELDAR: Nonlinear Waveform Inversion for Quantitative Ultrasound}


\maketitle

\begin{abstract}
Due to its non-invasive and non-radiating nature, along with its low cost, ultrasound (US) imaging is widely used in medical applications. Typical B-mode US images have limited resolution and contrast and weak physical interpretation. Inverse US methods were developed to reconstruct the media's speed-of-sound (SoS) based on a linear acoustic model. However, the wave propagation in medical US is governed by nonlinear acoustics, which introduces more complex behaviors neglected in the linear model. In this work we propose a nonlinear waveform inversion (NWI) approach for quantitative US, that considers a nonlinear acoustics model to simultaneously reconstruct multiple material properties, including the medium's SoS, density, attenuation, and nonlinearity parameter. We thus broaden current inverse US approaches, such as the full waveform inversion (FWI) algorithm, by considering nonlinear media, and additional physical parameters. We represent the nonlinear acoustic model by means of a recurrent neural network, which enables us to apply advanced optimization algorithms borrowed from the deep learning toolbox and achieve more efficient reconstructions compared to the FWI method. We evaluate the performance of our approach on in-silico data and show that neglecting nonlinear effects may result in substantial degradation in the reconstruction, paving the way of NWI into clinical applications.
\end{abstract}


\section{Introduction}
\label{sec:introduction}
Due to its non-invasive and non-ionizing nature, ultrasound (US) imaging is widely used in medical applications. In US imaging, an  image  is  generated  by  transmitting  a  series of  acoustic pulses from an array of transducer elements.  The transmitted  pulses  propagate  through  different  tissues,  leading  to  a sequence of reflections and refractions, which create echoes that are then detected by the same array. After acquiring the US signal, beamforming algorithms are used to align the signals from different transducer elements properly, and the data acquired from multiple transmission schemes are combined to generate a US image \cite{VanVeen1988Beamformig,Steinberg1992,Thomenius1996}.

Typical Brightness-mode (B-mode) images are generated by applying corresponding time delays to the acquired signal and averaging over the channels with tailored weights \cite{Viola2005,Synnevag2007,Holfort2007,Vignon2008,Synnevag2009,Yonina2019iMap}. Nevertheless, B-mode images do not necessarily provide sufficient contrast for certain anatomical structures and have poor physical interpretation.

Imaging of physical properties of the material such as speed-of-sound (SoS), density, acoustic attenuation, and elasticity, is known to have valuable differentiation capabilities and improve medical diagnosis \cite{Bamber1979, Szabo2013}. For example, SoS maps can discern between benign and malignant breast tumors \cite{Goss1978,Goss1980,Li2009,Sak2017,Ruby2019}, identify muscle loss and fatty muscular degeneration (sarcopenia) in seniors \cite{Sanabria2019}, and differentiate between healthy and diseased tissues such as in human and animal livers, affected by nonalcoholic fatty liver disease (NAFLD) \cite{Sehgal1986,Ghoshal2012}. Acoustic attenuation maps can improve the diagnosis of non-healthy tissues \cite{Bamber1979}. Finally, the tissue density can indicate a risk for breast cancer \cite{Wang2014}, and quantify the level of fat and steatosis in the liver, which is critical to monitor NAFLD and nonalcoholic steatohepatitis (NASH) \cite{Idilman2013,Tang2013}.

Inverse US algorithms seek to reconstruct the properties of a medium based on the acquired US signal. A standard method to solve the inverse US problem is the full waveform inversion (FWI) algorithm  \cite{Tarantola1984}, a computational technique initially developed in geophysics. It relies on a physical wave propagation model and therefore explains a broader range of phenomena compared to B-mode images. To reconstruct the properties, the algorithm utilizes an iterative gradient-based approach. The gradients are computationally heavy \cite{Virieux2009_2}, which increases the algorithm's time complexity and precludes real-time usage. Moreover, the FWI algorithm assumes a linear wave propagation model, which is inaccurate in medical setups. In medical US, the transmitted wave propagates in a highly nonlinear media \cite{Wells1999}.  As a result of neglecting nonlinear effects, some physical properties (such as the medium's nonlinearity), are not possible to extract using the FWI algorithm.
Furthermore, additional inverse algorithms that exploit the acoustic wave's reciprocity, cannot be applied, since nonlinear media do not satisfy time-reversal symmetry \cite{Blancharda2018}.

More efficient inverse US methods, use a differential path matrix to estimate the SoS of the medium \cite{Rau2021}. The differential path matrix links the SoS distribution to the obtained time delays at the transducer. However, the model only estimates the medium's SoS, since it depends on geometric considerations rather than the wave equation.
Alternative solutions to the inverse US problem, use machine learning approaches such as deep neural networks \cite{Jush2020}. Such approaches require a significant amount of medical data with known SoS maps for network training, and the obtained models have poor interpretability.

In this work, we propose the nonlinear waveform inversion (NWI) algorithm, a model-based recurrent neural network (RNN) approach to solve inverse nonlinear physical problems. We apply it to the inverse nonlinear US problem and recover the medium's properties. Since the algorithm is no longer constrained to the linear acoustic model, additional physical properties can be reconstructed, such as the medium's nonlinearity.

Nonlinearity in acoustics is exploited in a few applications, and enables to increase the amount of information extracted from an US scan. For example, harmonic imaging techniques, use the nonlinearity property to generate a second harmonic, which leads to substantial improvement in contrast and resolution of the beamformed images \cite{Averkiou2000}. However, to the best of our knowledge, there is no waveform inversion technique that exploits the nonlinearity, in order to improve the inversion, along with reconstructing additional properties.

The NWI algorithm is based on representing the wave equation as an RNN, with an architecture determined by the physical model. Contrary to the common usage of neural networks, we do not use the networks to learn a set of parameters but rather to express the wave equation by means of an RNN. This representation enables us to apply optimization algorithms used in the deep-learning field and paves the way to very efficient implementation. The generality of this approach allows reconstructing multiple properties simultaneously using a nonlinear acoustics (NLA) model, which better captures the wave's propagation in the human body \cite{Wells1999}. Moreover, the network representation enables us to achieve more efficient reconstruction compared to the FWI method, thus reducing the time and cost of using inverse techniques. 

To demonstrate the capabilities of the proposed approach, we reconstruct the properties of a simulated medium with characteristics corresponding to human tissues (fat and liver), and evaluate the NWI algorithm using the normalized root-mean-square-error (RMSE), as done in previous works on SoS reconstruction \cite{Bernhardt2020,Rau2021}. We compare the results with the FWI reconstructions. 
Although these simulations do not mimic human tissues, they serve as simpler examples to validate the theoretical findings. A more extensive analysis with clinically accurate simulations should be conducted to evaluate the contribution of NWI to clinical waveform inversion.
Using our approach, the reconstruction error is reduced by a factor up to 2.2, which emphasizes the importance of considering an adapted nonlinear model. Furthermore, the NWI algorithm enlarges the set of physical properties that can be reconstructed from the US signal compared to the FWI algorithm, such as the medium's nonlinearity. Finally, although the NLA model increases the problem's complexity, the NWI algorithm reduces the computational complexity compared to the FWI method.

The rest of the paper is organized as follows. In Section \ref{sec:Background}, we present the FWI algorithm which is the commonly used method today for solving the inverse US problem. In Section \ref{sec:method}, we derive the NWI algorithm and show how it generalizes the FWI technique. To capture nonlinear effects, our approach uses the discretized nonlinear acoustic wave equation, presented in Section \ref{sec:Discretization of the wave equation}. The NWI algorithm relies on the RNN representation of the wave equation described in Section \ref{sec:Recurrent neural network representation}, and is detailed in Section \ref{sec:Properties reconstruction}. We demonstrate the performance of NWI and compare it to FWI in Section \ref{sec:Results}. We summarize the key points of our work in Section \ref{sec:Discussion}.

Throughout the paper, we use boldface lower-case and upper-case for vectors and matrices, respectively. The vectorization, maximum, convolution, Frobenius norm, transpose, element-wise multiplication (Hadamard product), and Kronecker product operators are written as $\text{vec}(\cdot)$, $\max(\cdot)$, $*$, $\norm{\cdot}_F$, $(\cdot)^T$, $\odot$, and $\otimes$ respectively. Matrices division and exponentiation are performed element-wise. Finally, $\R$ denotes the set of real numbers. 

\section{Problem Formulation and Background}
\label{sec:Background}
We begin by formulating the inverse US problem, and then present the FWI algorithm \cite{Tarantola1984} which is commonly used today to solve it.

Propagation of acoustic waves in solids is determined by the physical properties of the material, such as the speed-of-sound ($c_0$), the density ($\rho_0$), the attenuation/damping ($D$), Young's modulus ($E$), Lam\'e constants ($\lambda, \mu$), and more \cite{Yao2018,Hamilton1998}. These properties depend on the spatial coordinates and can potentially also depend on the frequency of the applied pulse. They entirely determine the generated wave inside the medium.
In Section \ref{sec:Problem formulation}, we formulate the US inverse problem to reconstruct the medium's properties based on US data. In Section \ref{sec:Full Waveform Inversion Algorithm}, we introduce the FWI algorithm and discuss the assumptions it relies on.

\subsection{The Inverse Nonlinear US Problem}
\label{sec:Problem formulation}
Consider an US transducer array with $n_c$ elements, which insonifies the medium and measures the acoustic wave (pressure) on the elements, at $n_t$ time steps. The measured signal, $\mathbf{M} \in \R^{n_c \times n_t}$, is obtained by sampling the acquired acoustic wave with temporal intervals of $\Delta_t$ at the elements' locations. The US signal is generated from an acoustic pulse $\mathbf{F} \in \R^{n_x \times n_z \times n_t}$, where we discretize the spatial coordinates with an interval of $\Delta$, forming a grid of size $n_x \times n_z$. The acoustic pulse, $\mathbf{F}$, specifies the exterior force that is applied by the transducer's elements, as a function of time and space. Multiple types of acoustic pulses are commonly used, such as plane waves, focused beams, and diverging waves. A derivation of the acoustic pulse required for focused beams is presented in Appendix \ref{sec:appendix Acoustic Pulse}. The measurement of the acoustic wave is restricted to the transducer elements' locations. We denote by $\mathcal{R}(\cdot)$ the restriction operator, which given the pressure field over the $n_x \times n_z$ grid, extracts the pressure at the $n_c$ elements' locations. The restriction operator is defined by the shape of the transducer array. Various transducer types are regularly used, such as linear, convex and endocavitary probes.

The wave propagation is governed by a chosen physical model, with $n_\theta$ model parameters. We denote by $\left\{ \boldsymbol \Theta_i \right\}_{i = 1}^{n_\theta}$ the set of parameters, such that $\boldsymbol \Theta_i \in \R^{n_x \times n_z}$ for $i \in [1, n_\theta]$. The imaged material has ground-truth (GT) properties denoted by $\boldsymbol \Theta_i^{\star} \in \R^{n_x \times n_z}$ for $i \in [1, n_\theta]$. In this paper, we consider US signals for medical applications, where the medium is nonlinear (see Section \ref{sec:Discretization of the wave equation}), such that the measured signal, commonly referred to as channel-data in US terminology, can be written as 
\begin{equation}
\mathbf{M} = \mathcal{R}(\mathcal{N}(\left\{ \boldsymbol \Theta_i^{\star} \right\}_{i = 1}^{n_\theta}, \mathbf{F})) + \mathbf{N}
    \label{eq:f_function}
\end{equation}
where $\mathbf{N} \in \R^{n_c \times n_t}$ is an additive i.i.d. noise with zero mean. Here, $\mathcal{N}(\cdot)$ is the NLA operator, which given the medium's properties and the acoustic pulse, returns the acoustic wave. A detailed derivation of the NLA operator will be given in Section \ref{sec:Discretization of the wave equation}.

The goal of inverse US algorithms, is to estimate the medium's physical properties, based on the measured US signal, $\mathbf{M}$.
Formally, given a loss function, $L$, the goal is to find the properties that reproduce the experimental data, by solving the following minimization problem
\begin{equation}
    \left\{ \hat{\boldsymbol \Theta}_i \right\}_{i = 1}^{n_\theta} = \underset{\left\{ \boldsymbol \Theta_i \right\}_{i = 1}^{n_\theta}}{\arg\min} \ L(\left\{ \boldsymbol \Theta_i \right\}_{i = 1}^{n_\theta}, \mathbf{M}).
    \label{eq:minimization_problem}
\end{equation}
Usually, the additive noise is assumed to be normally distributed, and accordingly, the loss is chosen as the $L_2$ norm of the difference between the measured and predicted data.
The predicted data can be written as
\begin{equation}
\mathbf{P} = \mathcal{R}(\mathcal{N}(\left\{ \boldsymbol \Theta_i \right\}_{i = 1}^{n_\theta}, \mathbf{F}))
    \label{eq:pred_signal}
\end{equation}
where $\mathbf{P} \in \R^{n_c \times n_t}$.
The main difference between the predicted and measured signals, arise from the additive noise which cannot be approximated by a physical model.
Due to the added noise and the inherent complexity of the nonlinear optimization problem, multiple solutions are expected.

\subsection{Full Waveform Inversion Algorithm}
\label{sec:Full Waveform Inversion Algorithm}
The FWI method, is a computational approach to solve (\ref{eq:minimization_problem}) for the $L_2$ loss, under the assumption of a linear acoustic model \cite{Tarantola1984}.
To reconstruct the medium's properties, the FWI algorithm uses a gradient-based optimizer to minimize the loss, $L$, between the predicted and observed data:
\begin{equation}
    L = \frac{1}{2} \norm{\mathbf{p} - \mathbf{m}}_2^2
    \label{eq:misfit_FWI}
\end{equation}
where $\mathbf{m} = \text{vec}(\mathbf{M}) \in \R^{n_c n_t \times 1}$ is the measured (observed) data organized as a column vector, and $\mathbf{p} = \text{vec}(\mathbf{P}) \in \R^{n_c n_t \times 1}$ is the predicted data which depends on the estimated properties $\mathbf{p} = \mathbf{p}\left( \left\{ \boldsymbol \Theta_i \right\}_{i = 1}^{n_\theta} \right)$. As a result, also the loss depends on the material's properties. 
At each iteration, the iterative algorithm computes the derivatives of the loss with respect to the properties, which are used to update the estimations.
Denote by $\boldsymbol \theta_i = \text{vec}(\boldsymbol \Theta_i) \in \R^{n_x n_z \times 1}$ for $i \in [1,n_\theta]$, the medium's properties organized as column vectors. The derivative of the loss with respect to each of the material's properties, $\boldsymbol \theta_i$ for $i \in [1,n_\theta]$, is
\begin{equation}
    \frac{\partial L}{\partial \boldsymbol \theta_i} = \left( \frac{\partial \mathbf{p}}{\partial \boldsymbol \theta_i} \right)^T (\mathbf{p} - \mathbf{m}).
    \label{eq:misfit_FWI_derivative}
\end{equation}

The predicted data is a subset of the pressure wavefield $\mathbf{u} \in \R^{n_x n_z n_t \times 1}$, which represents the pressure as a function of time and space, organized as a column vector. The predicted data is obtained by applying the restriction operator on the pressure wavefield. In vector notation, the restriction operator, $\mathcal{R}$, can be expressed in matrix form as $\mathbf{R} \in \R^{n_c n_t \times n_x n_z n_t}$, such that:
\begin{equation}
    \mathbf{p} =\mathbf{R} \mathbf{u}.
    \label{eq:restriction_matrix}
\end{equation}

The pressure wavefield $\mathbf{u}$, is obtained by solving the wave equation. The FWI algorithm assumes a linear wave equation:
\begin{equation}
    \mathbf{A}\mathbf{u} = \mathbf{f}
    \label{eq:linear_model_vector}
\end{equation}
 where $\mathbf{f} = \text{vec}(\mathbf{F}) \in \R^{n_x n_z n_t \times 1}$ is the applied acoustic pulse, organized as a column vector. In addition, $\mathbf{A} \in \R^{n_x n_z n_t \times n_x n_z n_t}$ is the linear wave equation operator written in discrete form. This operator, governs the wave propagation, and can be computed explicitly from the set of physical properties: $\mathbf{A} = \mathbf{A}\left( \left\{ \boldsymbol \Theta_i \right\}_{i = 1}^{n_\theta} \right)$. As an example, we derive the linear wave equation operator explicitly, for homogeneous media, in Appendix \ref{sec:appendix Linear acoustic operator}. The pressure wavefield, $\mathbf{u}$, is obtained by solving (\ref{eq:linear_model_vector}). 
 For a given set of parameters, the wave equation (\ref{eq:linear_model_vector}) has a unique solution, and since $\mathbf{A}$ is a square matrix, it is invertible, leading to
 \begin{equation}
    \mathbf{u} = \mathbf{A}^{-1} \mathbf{f}.
    \label{eq:explicit_pressure_linear}
\end{equation}

We note that $\mathbf{A}$ and $\mathbf{u}$ depend on the material's properties, whereas $\mathbf{f}$ does not. Taking the derivative of (\ref{eq:linear_model_vector}), leads to
\begin{equation}
    \mathbf{A}\frac{\partial \mathbf{u}}{\partial \boldsymbol \theta_i} + \frac{\partial \mathbf{A}}{\partial \boldsymbol \theta_i} \mathbf{u} = 0.
    \label{eq:derivative_linear_model}
\end{equation}
The term $\frac{\partial \mathbf{A}}{\partial \boldsymbol \theta_i} \in \R^{n_x n_z n_t \times n_x n_z n_t \times n_x n_z}$, is a three dimensional matrix, such that $\left( \frac{\partial \mathbf{A}}{\partial \boldsymbol \theta_i} \mathbf{u} \right) \in \R^{n_x n_z n_t \times n_x n_z}$. Since $\mathbf{A}$ is invertible, we can write
\begin{equation}
    \frac{\partial \mathbf{u}}{\partial \boldsymbol \theta_i} = - \mathbf{A}^{-1} \left(\frac{\partial \mathbf{A}}{\partial \boldsymbol \theta_i} \mathbf{u} \right).
    \label{eq:derivative_u}
\end{equation}
Noting that the restriction matrix, $\mathbf{R}$, is independent of the physical properties, we have for all $i \in [1,n_\theta]$:
\begin{equation}
    \frac{\partial \mathbf{p}}{\partial \boldsymbol \theta_i} = \mathbf{R} \frac{\partial \mathbf{u}}{\partial \boldsymbol \theta_i} = - \mathbf{R} \mathbf{A}^{-1} \left(\frac{\partial \mathbf{A}}{\partial \boldsymbol \theta_i} \mathbf{u} \right).
    \label{eq:derivative_p}
\end{equation}
Substituting (\ref{eq:derivative_p}) into (\ref{eq:misfit_FWI_derivative}) gives  
\begin{equation}
    \frac{\partial L}{\partial \boldsymbol \theta_i} = - \left( \frac{\partial \mathbf{A}}{\partial \boldsymbol \theta_i} \mathbf{u} \right)^T \mathbf{A}^{-T} \mathbf{R}^T (\mathbf{p} - \mathbf{m}).
    \label{eq:derivative_L}
\end{equation}
The term $\mathbf{r} = \mathbf{A}^{-T} \mathbf{R}^T (\mathbf{p} - \mathbf{m})$ can be thought of as injecting the residual signal $(\mathbf{p} - \mathbf{m})$ into the medium through the transducer array, and propagating the wave backward in time with $\mathbf{A}^{-T}$. The method is summarized in Algorithm \ref{alg:algorithm_FWI}.
\begin{algorithm}[ht!]
\begin{algorithmic}
\caption{Full waveform inversion}
\label{alg:algorithm_FWI}
\Statex \textbf{Inputs:} 
\Statex \ \ \ \ $\mathbf{m} \in \R^{n_c  n_t \times 1}$ \Comment{The measured channel data}
\Statex \ \ \ \ $\mathbf{f} \in \R^{n_x  n_z  n_t \times 1}$ \Comment{The applied pulse}
\Statex \ \ \ \ $\boldsymbol \theta_i^{(0)} \in \R^{n_x  n_z \times 1}, \forall i \in [1,n_\theta]$ \Comment{Set of initial properties}
\State $\boldsymbol \theta_i \xleftarrow[]{} \boldsymbol \theta_i^{(0)}, \forall i \in [1,n_\theta]$ \Comment{Properties' initialization}
\State \textbf{while} stopping criteria is not satisfied \textbf{do}
\State \ \ \ \  Compute the pressure field $\mathbf{u} = \mathbf{A}^{-1} \mathbf{f}$
\State \ \ \ \  Compute the predicted signal $\mathbf{p} = \mathbf{R} \mathbf{u}$
\State  \ \ \ \ Calculate the gradients for $\boldsymbol \theta_i, \forall i \in [1,n_\theta]$: 
\State  \ \ \ \ \ \ \ \ \ \ \ \ \ \ \ \ \  $\frac{\partial L}{\partial \boldsymbol \theta_i} = - \left( \frac{\partial \mathbf{A}}{\partial \boldsymbol \theta_i} \mathbf{u} \right)^T \mathbf{A}^{-T} \mathbf{R}^T (\mathbf{p} - \mathbf{m})$
\State  \ \ \ \ Update the estimators $\boldsymbol \theta_i, \forall i \in [1,n_\theta]$ using a gradient-based optimizer
\State \textbf{end while}
\Statex \textbf{Output:} $\left\{ \boldsymbol \theta_i \right\}_{i = 1}^{n_\theta}$  \Comment{The set of reconstructed properties}
\end{algorithmic}
\end{algorithm}

The computational complexity of each iteration is determined by the complexity of the gradients computation. Computing the gradient with respect to single element in $\boldsymbol \theta_i$, requires a applying the derivative of $\mathbf{A}$ with respect to that element on $\mathbf{u}$, and multiplying the result by the residual wave, $\mathbf{r}$. The computational complexity of these operations is $\O(n_x  n_z  n_t)$, as detailed in \cite{Virieux2009_2}. This process is repeated for each element in $\boldsymbol \theta_i$, leading to an overall computational complexity of $\O((n_x  n_z)^2  n_t)$.

Algorithm \ref{alg:algorithm_FWI} suffers from a few inherent drawbacks. First, the algorithm considers only linear wave equations, which is not suited for medical applications, where higher orders of $\mathbf{u}$ affect the wave's propagation. Neglecting nonlinear effects, impairs the reconstruction and prevents estimating some properties, such as the medium's nonlinearity. Second, due to the matrix representation of the wave, the gradient computation requires high computational complexity, which precludes usage of this method in many practical settings. 

\section{Nonlinear Waveform Inversion}
\label{sec:method}
In this section, we introduce the NWI algorithm, which is a model-based RNN approach for properties estimation. The method generalizes the FWI algorithm by considering nonlinear physical models, which are imperative for medical setups.  The NWI algorithm broadens the set of properties that can be estimated, and considers nonlinear effects in the reconstruction process. The algorithm exploits the neural network representation of the wave equation, and although the nonlinear models are more complex, it achieves more efficient gradients' computation. 

In Section \ref{sec:Discretization of the wave equation}, we present the discrete NLA model adopted from \cite{Yao2018}, used to numerically solve the inverse US problem. We then show in Section \ref{sec:Recurrent neural network representation}, that the acoustic wave can be represented as an RNN, enabling us to apply advanced optimization algorithms developed for deep neural networks, in order to solve the inverse US problem in (\ref{eq:minimization_problem}), more efficiently than FWI. In Section \ref{sec:Properties reconstruction} we present our proposed reconstruction technique that considers nonlinear effects in the estimation process.  Finally, in Section \ref{sec:multiple pulses scenario} we present a generalized version of the algorithm adapted for practical settings. In practice, multiple insonifications are created by the US transducer, generating a series of measured channel-data. In Algorithm \ref{alg:algorithm_multiple_pulses}, we show how to combine the information from multiple measurements, to improve the reconstruction. 

\subsection{Nonlinear Acoustic Wave Equation}
\label{sec:Discretization of the wave equation}
Nonlinearity in acoustic waves arises from coupling between the material's properties and the thermodynamics fields, resulting in additional frequency components introduced to the system. When traveling inside the material, the wave alters various thermodynamics fields (such as the temperature and density) which affect the propagation, leading to self distortion.

In this work, we consider the nonlinear acoustic wave equation in two dimensions, given by the lossy Westervelt equation \cite{Yao2018,Hamilton1998}:
\begin{equation}
\begin{aligned}
    & \frac{\partial^2 u}{\partial t^2} + 2 D \frac{\partial u}{\partial t} + D^2 u + \frac{\beta}{c_0^2 \rho_0} \frac{\partial^2 u^2}{\partial t^2} \\
    & = c_0^2 \rho_0 \left( \frac{\partial}{\partial x}\left(\frac{1}{\rho_0} \frac{\partial u}{\partial x}\right) + \frac{\partial}{\partial z}\left(\frac{1}{\rho_0} \frac{\partial u}{\partial z}\right) \right) + F
    \label{eq:acoustic_wave_2D}
\end{aligned}
\end{equation}
where $F$ is the continuous acoustic pulse, $c_0, \rho_0, D$, and $\beta$ are the medium's continuous SoS, density, attenuation, and nonlinearity parameter, respectively, and $u$ is the resulting acoustic wave. In medical applications, the imaged materials are highly nonlinear: blood ($\beta = 4.0$), fat ($\beta = 6.0$), and muscle ($\beta = 4.7$) \cite{Wells1999}. We note that (\ref{eq:acoustic_wave_2D}) is nonlinear in $u$, and therefore this model cannot be used by the FWI algorithm. As a result, the nonlinearity parameter, $\beta$, cannot be estimated using the FWI approach.

In order to obtain a discrete version of the lossy Westervelt equation, the wave is sampled on a discrete grid.
The discrete form of the temporal derivative, is given by a weighted average of past time samples, and the discrete spatial derivative of a signal is obtained by filtering (convolving) it with the discrete gradient or Laplacian filters. We denote by $\mathbf{U} \in \R^{n_x \times n_z \times n_t}$ the pressure as a function of time and space, in matrix form. 
Following the derivation in \cite{Yao2018}, the discrete wave is
\begin{equation}
\begin{aligned}
    & \frac{1}{\Delta_t^2} \left( \mathbf{1} + 2\frac{\mathbf{B} \odot \mathbf{U}[n - 1]}{\mathbf{C}^2 \odot \mathbf{Q}} \right) \odot \left( \mathbf{U}[n] -2\mathbf{U}[n - 1] + \mathbf{U}[n - 2] \right) \\
    & + \frac{2}{\Delta_t^2} \frac{\mathbf{B}}{\mathbf{C}^2 \odot \mathbf{Q}} \odot \left( \mathbf{U}[n - 1] - \mathbf{U}[n - 2]\right)^2 \\ 
    & + \frac{2}{\Delta_t} \mathbf{D} \odot \left( \mathbf{U}[n - 1] - \mathbf{U}[n - 2] \right) + \mathbf{D}^2 \odot \mathbf{U}[n - 1] \\
    & \ \ \ \ \ \ \ \ \ \ \ \ \ \ = \mathbf{C}^2 \odot \mathbf{Q} \left( \nabla_D * \left( \frac{1}{\mathbf{Q}} \right) \right) \cdot \left( \nabla_D * \mathbf{U}[n - 1] \right) \\
    & \ \ \ \ \ \ \ \ \ \ \ \ \ \ + \mathbf{C}^2 \odot \mathbf{Q} \left(\nabla_D^2 * \mathbf{U}[n - 1] \right) + \mathbf{F}[n]
\label{eq:discrete_acoustic_wave_2D}
\end{aligned}
\end{equation}
where $\mathbf{U}[n'], \mathbf{F}[n'] \in R^{n_x \times n_z}$ are the acoustic wave and applied pulse at the $n'$th time step, $\mathbf{C}, \mathbf{Q}, \mathbf{D}, \mathbf{B} \in R^{n_x \times n_z}$ are the discrete SoS, density, attenuation, and nonlinearity parameter, respectively, $\nabla_D$ is the discrete gradient filter, $\nabla^2_D$ is the discrete Laplacian filter, and $\mathbf{1} \in \R^{n_x \times n_z}$ is a matrix of ones. All matrix divisions and exponentiation are performed element-wise. In this work, we aim to recover the following properties $\left\{ \boldsymbol \Theta_i \right\}_{i = 1}^{n_\theta = 4} = \{ \mathbf{C}, \mathbf{Q}, \mathbf{D}, \mathbf{B} \}$. However, additional physical properties can be reconstructed, by choosing an appropriate model.

To ensure the convergence of the numerical equation to a valid partial differential equation (PDE) solution, we enforce the Courant-Friedrichs-Lewy (CFL) condition which imposes restrictions on the relation between the spatial and temporal intervals - $\Delta$ and $ \Delta_t$ \cite{CFL1928}. Specifically, we require that $C_r = \max(\mathbf{C})  \frac{\Delta_t}{\Delta} \leq 1$, where $C_r$ is the Courant number.

The wave field $\mathbf{U}$ is the solution to (\ref{eq:discrete_acoustic_wave_2D}), and depends on the set of physical properties. It can be expressed as
\begin{equation}
    \mathbf{U} = \mathcal{N}\left( \left\{ \boldsymbol \Theta_i \right\}_{i = 1}^{n_\theta}, \mathbf{F} \right)
    \label{eq:nonlinear_acoustic_operator}
\end{equation}
where $\mathcal{N}$ is the NLA operator. 
Similarly to the FWI algorithm, to reconstruct the material's properties, we will compute the gradients of the loss with respect to the properties - $\frac{\partial L}{\partial \boldsymbol \Theta_i}$ for $i \in [1,n_\theta]$. As a result, the computation of the gradients of the acoustic wave with respect to the properties, $\frac{\partial \mathbf{U}}{\partial \boldsymbol \Theta_i}$ for $i \in [1,n_\theta]$, plays an important role in the reconstruction process.
In Section \ref{sec:Recurrent neural network representation}, we derive a convenient way to represent the NLA operator, by means of a recurrent neural network, which enables efficient gradients computation.

\subsection{Recurrent Neural Network Representation}
\label{sec:Recurrent neural network representation}
RNNs are a class of artificial neural networks where the nodes are connected over a temporal sequence, such that the same operation is applied at each time step, allowing it to exhibit dynamic behavior.

To compute the discrete wave efficiently, we represent it as an RNN, where each wave equation's time step coincides with the corresponding RNN's time step. This is done by rearranging (\ref{eq:discrete_acoustic_wave_2D}), and obtaining the recurrence relation between $\mathbf{U}[n]$ and its past time samples:
\begin{equation}
\begin{aligned}
    \mathbf{U}[n] = & \frac{\mathbf{G}_2}{\mathbf{G}_1} \odot \mathbf{U}[n - 1] + \frac{\mathbf{G}_3}{\mathbf{G}_1} \odot \mathbf{U}[n - 2] &\\
    + & \frac{\mathbf{G}_4}{\mathbf{G}_1} \odot \left(\mathbf{U}[n - 1] - \mathbf{U}[n - 2] \right)^2 &\\
    + & \frac{\mathbf{C}^2 \odot \mathbf{Q}}{\mathbf{G}_1} \odot \left( \nabla_D * \left( \left( \frac{1}{\mathbf{Q}} \right) \right) \cdot \nabla_D * \mathbf{U}[n - 1] \right) &\\
    + & \frac{\mathbf{C}^2 \odot \mathbf{Q}}{\mathbf{G}_1} \odot \nabla_D^2 * \mathbf{U}[n - 1] &\\
    + & \frac{1}{\mathbf{G}_1} \mathbf{F}[n]&
\label{eq:DNN_representation}
\end{aligned}
\end{equation}
where
\begin{equation}
\begin{aligned}
    \mathbf{G}_1 = & \frac{1}{\Delta_t^2} \left( \mathbf{1} + 2\frac{\mathbf{B} \odot \mathbf{U}[n - 1]}{\mathbf{C}^2 \odot \mathbf{Q}} \right) \\
    \mathbf{G}_2 = & 2\mathbf{G}_1 - \mathbf{D}^2 + \frac{2}{\Delta_t}\mathbf{D} \\
    \mathbf{G}_3 = & -\mathbf{G}_1 + \frac{2}{\Delta_t} \mathbf{D} \\
    \mathbf{G}_4 = & -\frac{2}{\Delta_t^2} \frac{\mathbf{B}}{\mathbf{C}^2 \odot \mathbf{Q}}.
\label{eq:coeficients_a_DNN}
\end{aligned}
\end{equation}

\begin{figure}[t!]
    \centering
    \includegraphics[width=\columnwidth]{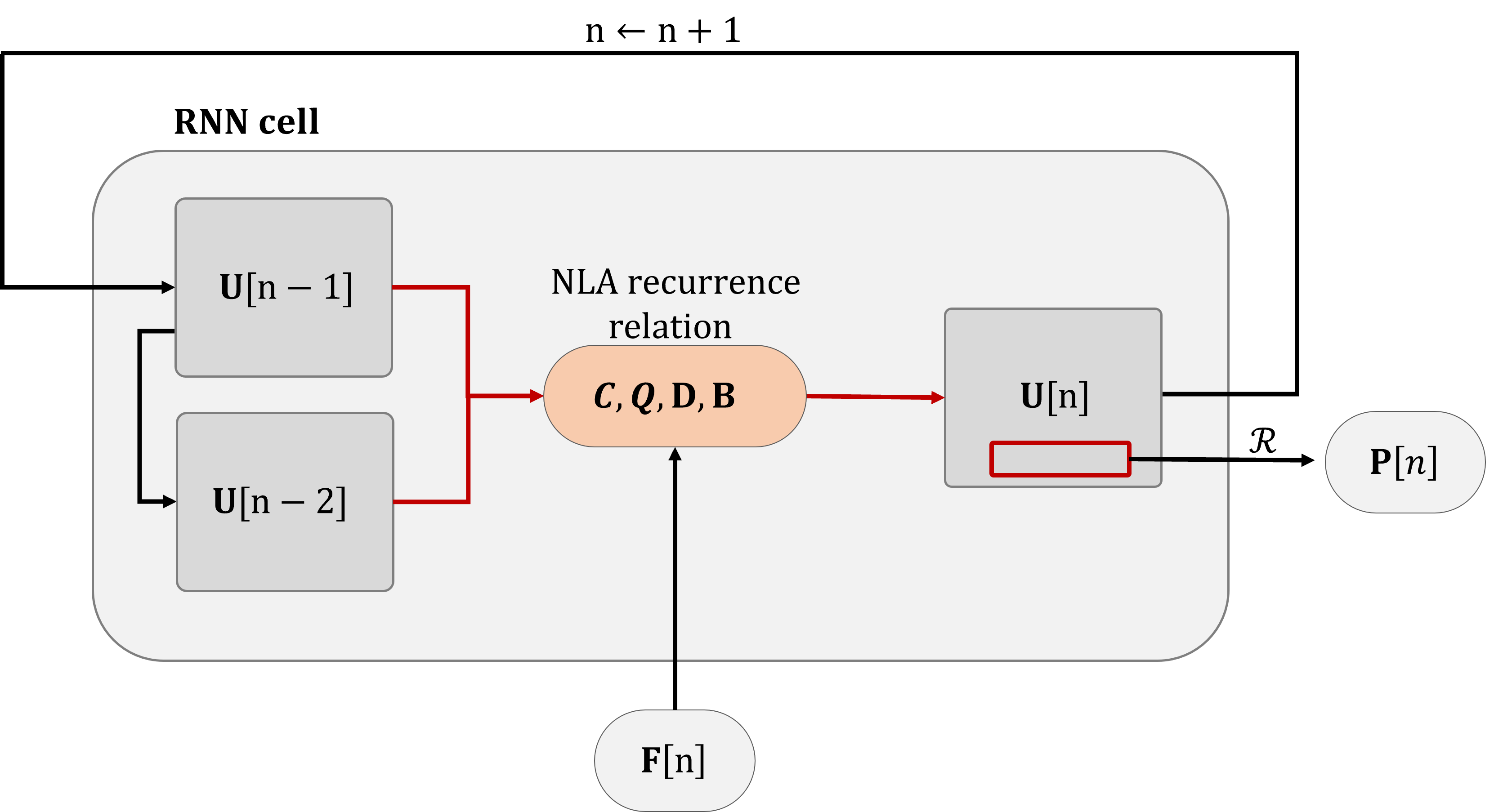}
    \caption{RNN representation of the nonlinear acoustic wave equation. At each iteration, the recurrent network computes the pressure at the next time step, depending on the medium's properties and past time samples, according to the NLA recurrent relation (\ref{eq:DNN_representation}). The pressure map $\mathbf{U}[n]$, is related to $\mathbf{U}[n - 2]$ and $\mathbf{U}[n - 1]$ by convolution operations with known kernels determined by the NLA wave equation (\ref{eq:DNN_representation}). The pulse applied by the transducer array, $\mathbf{F}$, serves as the input to the network at each iteration. The measured signal, $\mathbf{P}$, is obtained by restricting the acquired signal to the transducer's elements locations, using the restriction operator $\mathcal{R}$.}
    \label{fig:CNN_representation}
\end{figure}

The RNN performs the above operation repeatedly, where at each iteration, the network outputs the acoustic wave restricted to the array elements' locations, as illustrated in Fig. \ref{fig:CNN_representation}. We note that the recurrent relation is composed of element-wise multiplications and convolutions with kernels determined by the physical model (the gradient and Laplacian kernels). 

This representation will help us in Section \ref{sec:Properties reconstruction}, where we exploit the recurrence relation to backpropagate the gradients through the network.
Applying the backpropagation algorithm, on the RNN specified in (\ref{eq:DNN_representation}), returns the gradients of the wave with respect to the physical properties, $\frac{\partial \mathbf{U}}{\partial \boldsymbol \Theta_i}$ for $i \in [1,n_\theta]$. These gradients are essential in the reconstruction process, as they serve for computing $\frac{\partial \mathbf{P}}{\partial \boldsymbol \Theta_i}$ and $\frac{\partial L}{\partial \boldsymbol \Theta_i}$. In contrast to FWI, the gradients of any wave equation, can be computed based on this representation.

\subsection{Properties Reconstruction}
\label{sec:Properties reconstruction}
To reconstruct the medium's properties, we solve (\ref{eq:minimization_problem}) using a gradient-based iterative algorithm, such as gradient-descent, L-BFGS \cite{LBFGS1980}, Adam \cite{Adam2014}, and AdaDelta \cite{Adadelta2012}.
The gradients of the loss with respect to the physical properties, are computed with the backpropagation algorithm on the specified RNN (\ref{eq:DNN_representation}).

The backpropagation algorithm exploits the fact that the neural network is composed from sequences of basic computational steps, in order to compute the derivative of the output with respect to the inputs \cite{Rumelhart1986_Backprop}. The derivative computation consists of computing the derivatives of the basic operations on the network’s nodes, and propagating them through the network according to the chain rule. The implementation is done through automatic differentiation tools, which offer a general framework to construct neural networks and compute their derivatives through the backpropagation algorithm \cite{NEURIPS2019_9015}.

The set of estimated properties, $\left\{ \boldsymbol \Theta_i \right\}_{i = 1}^{n_\theta}$, is updated at each iteration of the iterative algorithm. For example, to estimate $\boldsymbol \Theta_i, \forall i \in [1,n_\theta]$ using gradient-descent, the algorithm will perform the following step at each iteration
\begin{equation}
    \boldsymbol \Theta_i \xleftarrow[]{} \boldsymbol \Theta_i - \alpha \frac{\partial L}{\partial \boldsymbol \Theta_i}
\label{eq:GD_example}
\end{equation}
where $\alpha$ is the learning rate, and $\frac{\partial L}{\partial \boldsymbol \Theta_i}$ is the derivative of the loss with respect to $\boldsymbol \Theta_i$, obtained by applying the backpropagation algorithm on the RNN (\ref{eq:DNN_representation}). The method is summarized in Algorithm \ref{alg:algorithm_RNN}.
\begin{algorithm}[ht!]
\begin{algorithmic}
\caption{Nonlinear waveform inversion}
\label{alg:algorithm_RNN}
\Statex \textbf{Inputs:} 
\Statex \ \ \ \ $\mathbf{M} \in \R^{n_c \times n_t}$ \Comment{The measured channel data}
\Statex \ \ \ \ $\mathbf{F} \in \R^{n_x \times n_z \times n_t}$ \Comment{The applied pulse}
\Statex \ \ \ \ $\boldsymbol \Theta_i^{(0)} \in \R^{n_x \times n_z}, \forall i \in [1,n_\theta]$ \Comment{Set of initial properties}
\State $\boldsymbol \Theta_i \xleftarrow[]{} \boldsymbol \Theta_i^{(0)}, \forall i \in [1,n_\theta]$ \Comment{Properties' initialization}
\State \textbf{while} stopping criteria is not satisfied \textbf{do}
\State  \ \ \ \ Backpropogate the gradients through the network: $\frac{\partial L}{\partial \boldsymbol \Theta_i}$ for $i \in [1,n_\theta]$
\State  \ \ \ \ Update the estimators $\boldsymbol \Theta_i$ for $i \in [1,n_\theta]$, according to the chosen optimizer
\State \textbf{end while}
\Statex \textbf{Output:} $\left\{ \boldsymbol \Theta_i \right\}_{i = 1}^{n_\theta}$  \Comment{The set of reconstructed properties}
\end{algorithmic}
\end{algorithm}

Contrary to the common use of neural networks, here, we do not perform learning, but rather express the objective function in terms of an RNN, in order to facilitate the gradients' computation. Once the acoustic wave is represented by an RNN, we can apply advanced optimization algorithms borrowed from the deep-learning toolbox, based on the backpropagation algorithm.

In contrast to the FWI algorithm, the RNN representation does not rely on the linearity of the model, and allows to capture more complex physical behaviors. Although we have demonstrated that this approach holds specifically for the NLA wave equation, this is part of a broader concept, which can be applied to additional physical problems such as in photoacoustics and seismology.  

Finally,  similarly to the FWI algorithm and contrary to geometry-based methods \cite{Rau2021}, the NWI approach can be applied with an arbitrary type of pulse $\mathbf{F}$ (plane wave, diverging wave, and focused beam), and transducer array geometry, captured by the restriction operator $\mathcal{R}$ (linear, convex, and endocavitary probes). 

\subsection{Complexity Analysis}
The algorithm's time complexity is crucial in practice, in particular for real-time imaging applications.
The time complexity of the forward and backward passes through the RNN, are obtained by performing $n_t$ convolutions of $n_x \times n_z$ matrices. This leads to a complexity of $\O(n_x  n_z  n_t)$, resulting in a complexity reduction by a factor of $n_x n_z$ compared to the FWI algorithm, despite the increase in the model's complexity due to nonlinearity.

To further reduce the size of the spatial grid and improve the efficiency of the algorithm, the acoustic wave can be restricted to the region of interest in the medium. However, naive boundary conditions will introduce further reflections to the system. To inhibit those unwanted reflections without increasing the computational complexity, perfectly matched layers (PMLs) are used \cite{Yao2018}. The PML applies a gradual attenuation of the acoustic wave near the simulation grid's boundaries, thus suppressing artificial reflections and mimicking a grid without boundaries.  
In our method, we add an artificial attenuation at the boundary of the grid, of the form:
\begin{equation}
    D(l) \propto \left( \frac{l}{L_P} \right)^2
    \label{eq:PML_eq}
\end{equation}
where $L_P$ is the width of the absorbing layer, and $l \in [0,L_P]$ denotes the distance to the boundary. 

\subsection{Reconstruction From Multiple Pulses}
\label{sec:multiple pulses scenario}
Until now, we discussed how to reconstruct the medium's properties, from an US signal generated by a single acoustic pulse. However, in practical settings, the US transducer performs a sequence of multiple insonifications, originating from multiple pulses. The pulses usually differ by their position and orientation inside the medium. The data collected from multiple insonifications can be combined to improve the reconstruction. 

We consider a sequence of $n_l$ insonifications, originating from pulses $\mathbf{F}_l \in \R^{n_x \times n_z \times n_t}$ for $l \in [1, n_l]$. As a result, a series of measurement data is acquired denoted by $\mathbf{M}_l \in \R^{n_c \times n_t}$ for $l \in [1, n_l]$:
\begin{equation}
\mathbf{M}_l = \mathcal{R}(\mathcal{N}(\left\{ \boldsymbol \Theta_i^{\star} \right\}_{i = 1}^{n_\theta}, \mathbf{F}_l)) + \mathbf{N}_l
    \label{eq:f_function_l}
\end{equation}
where $\mathbf{N}_l \in \R^{n_c \times n_t}, \forall l \in [1, n_l]$ are additive i.i.d. noises with zero mean.

To reconstruct the medium's properties, we use a local gradient-descent based algorithm \cite{Khaled2020}. This is an iterative algorithm, where at each iteration, we apply Algorithm \ref{alg:algorithm_RNN} on each of the acquired signals, and the resulting estimators are averaged to obtain the final estimation. Due to the locality of the method, the uses of Algorithm \ref{alg:algorithm_RNN} are independent and can be distributed over multiple processing units, using parallel computing tools. As a result, for $n_l$ processing units, the time complexity of each iteration is similar to Algorithm \ref{alg:algorithm_RNN}. Moreover, this algorithm is efficient in terms of communication between the processing units \cite{Khaled2020}, since the number of synchronization across all units is reduced compared to common gradient-descent based algorithms. The method is summarized in Algorithm \ref{alg:algorithm_multiple_pulses}.

\begin{algorithm}[ht!]
\begin{algorithmic}
\caption{Reconstruction from multiple insonifications}
\label{alg:algorithm_multiple_pulses}
\Statex \textbf{Inputs:} 
\Statex \ \ \ \ $\mathbf{M}_l \in \R^{n_c \times n_t}$, $\forall l \in [1,n_l]$ \Comment{The measured channel data}
\Statex \ \ \ \ $\mathbf{F}_l \in \R^{n_x \times n_z \times n_t}$, $\forall l \in [1,n_l]$ \Comment{The applied pulses}
\Statex \ \ \ \ $\boldsymbol \Theta_i^{(0)} \R^{n_x \times n_z}, \forall i \in [1,n_\theta]$ \Comment{Set of initial properties}
\State $\boldsymbol \Theta_i \xleftarrow[]{} \boldsymbol \Theta_i^{(0)}, \forall i \in [1,n_\theta]$ \Comment{Properties' initialization}
\State \textbf{for} number of iterations \textbf{do}
\State  \ \ \ Apply Algorithm \ref{alg:algorithm_RNN} in parallel $\forall l \in [1,n_l]$: 
\State  \ \ \ \ \ \  $\left\{ \boldsymbol \Theta_i^{(l)} \right\}_{i = 1}^{n_\theta} \xleftarrow[]{}$ Algorithm \ref{alg:algorithm_RNN} $(\mathbf{M}_l, \mathbf{F}_l, \left\{ \boldsymbol \Theta_i \right\}_{i = 1}^{n_\theta})$
\State  \ \ \ Average the estimators:
\State  \ \ \ \ \ \ \  $\boldsymbol \Theta_i \xleftarrow[]{} \frac{1}{n_l} \sum_{l=1}^{n_l} 
\boldsymbol \Theta_i^{(l)}$, $\forall i \in [1,n_\theta]$
\State \textbf{end for}
\Statex \textbf{Output:} $\left\{ \boldsymbol \Theta_i \right\}_{i = 1}^{n_\theta}$  \Comment{The set of reconstructed properties}
\end{algorithmic}
\end{algorithm}

\section{Simulation Results}
\label{sec:Results}
To test our method, we applied the NWI algorithm on in-silico data with known properties. In Section \ref{sec:Simulation setup}, we describe the settings of the US transducer along with the setup for the NWI algorithm. In Section \ref{sec:reconstructions}, we describe the simulation experiments we performed, following previous inverse US works for medical application \cite{Bernhardt2020,Rau2021}. We present the results obtained from the experiments with NWI, evaluate them, and compare them to the FWI reconstructions.

\subsection{Simulation Setup}
\label{sec:Simulation setup}
To demonstrate the capabilities of our approach, we reconstructed the properties of a $50 [mm] \times 50 [mm]$ simulated medium with similar characteristics as human tissues \cite{Duck1990}.  We used a linear transducer array with 80 elements. At each lateral emission, 16 transducer's elements are employed to generate an acoustic pulse $\mathbf{F}$, with a central frequency of $f_0 = 4 \ MHz$. The pulses are focused beams, with a focus depth of $5 \ [mm]$, and no steering angle. A more detailed explanation on focused beams can be found in Appendix \ref{sec:appendix Acoustic Pulse}. The transducer performs $n_l = 16$ consecutive lateral emissions of the focused beam, moving along the array with a stride of four elements between the emissions. 

The additive noises are normally distributed with zero means. The variance is chosen such that the obtain signal-to-noise ratio (SNR) is 20, imitating in-vivo US scans \cite{Benzarti2013}. Accordingly, we used the regularized $L_2$ loss:
\begin{equation}
\begin{aligned}
    L(\left\{ \boldsymbol \Theta_i \right\}_{i = 1}^{n_\theta}, \mathbf{m}) & = \norm{\mathbf{P}(\left\{ \boldsymbol \Theta_i \right\}_{i = 1}^{n_\theta}, \mathbf{F}) - \mathbf{M}}_F &\\
    & + \sum_{i = 1}^{n_\theta = 4}\lambda_{\boldsymbol \Theta_i} \norm{\mathcal{D}_S \boldsymbol \Theta_i}_F &
\label{eq:L2_loss}
\end{aligned}
\end{equation}
where $\mathbf{P}$ is the predicted signal (\ref{eq:pred_signal}), and $\mathbf{M}$ is the measured signal (\ref{eq:f_function}). Here, $\mathcal{D}_S$ is the Sobel regularization operator that enforces soft edges, and $\lambda_{\boldsymbol \Theta_i} \in \R$ for $i \in [1,n_\theta]$, control the level of regularization.
Additional regularizations can be employed, depending on the available prior knowledge on the reconstructed medium.
At each iteration, the estimations are updated with the Adam optimizer.

\begin{figure}[t!]
    \centering
    \includegraphics[width=\columnwidth]{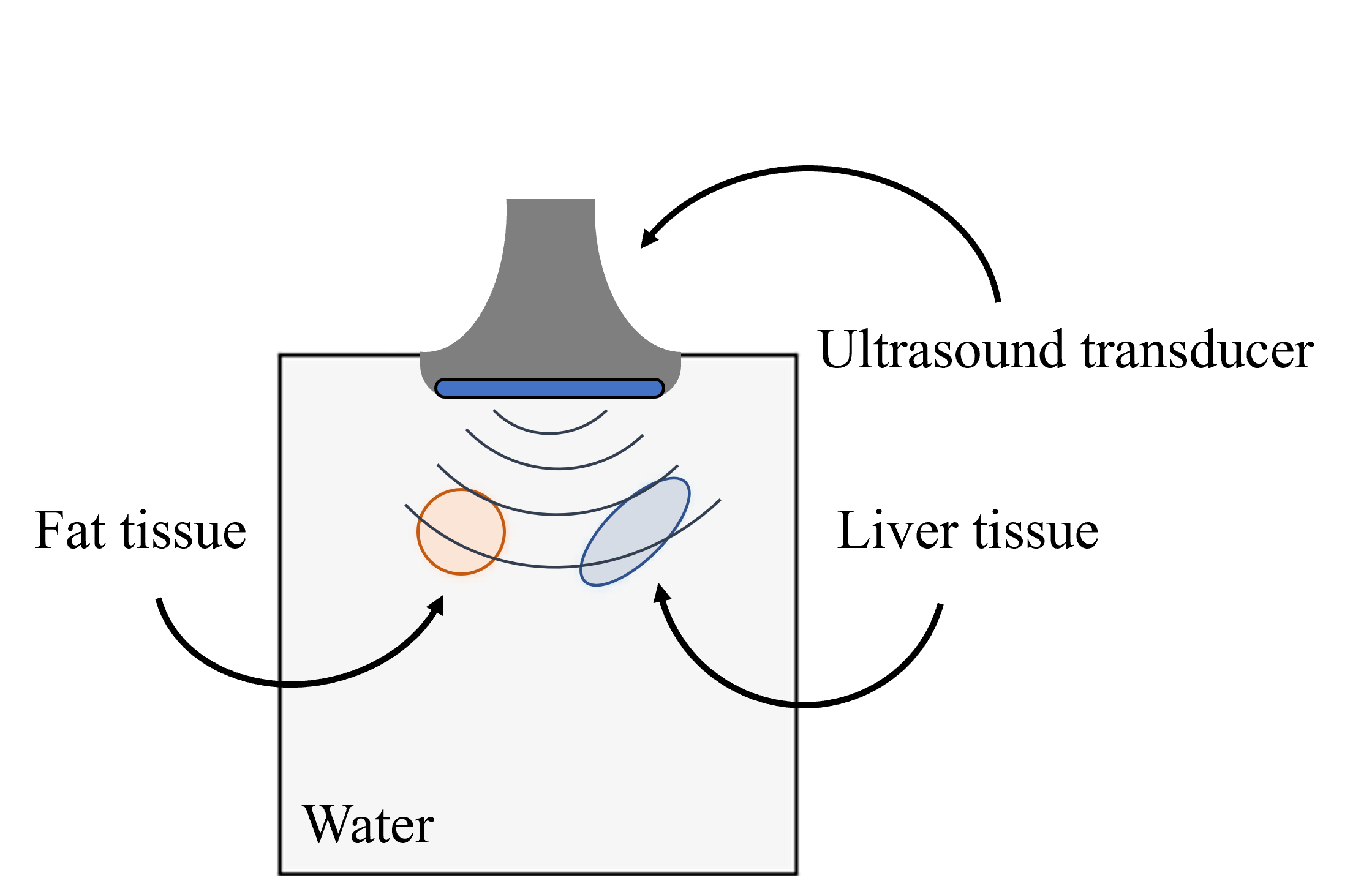}
    \caption{Illustration of the simulated medium. Water is represented in white. The left and right objects are fat and liver, respectively. The tissues and the water have different SoS, density, attenuation, and nonlinearity. The transducer array is located at the top.}
    \label{fig:Illustrarion}
\end{figure}

\subsection{Reconstructions}
\label{sec:reconstructions}

\begin{figure*}[t!]
    \centering
    \includegraphics[width=\textwidth]{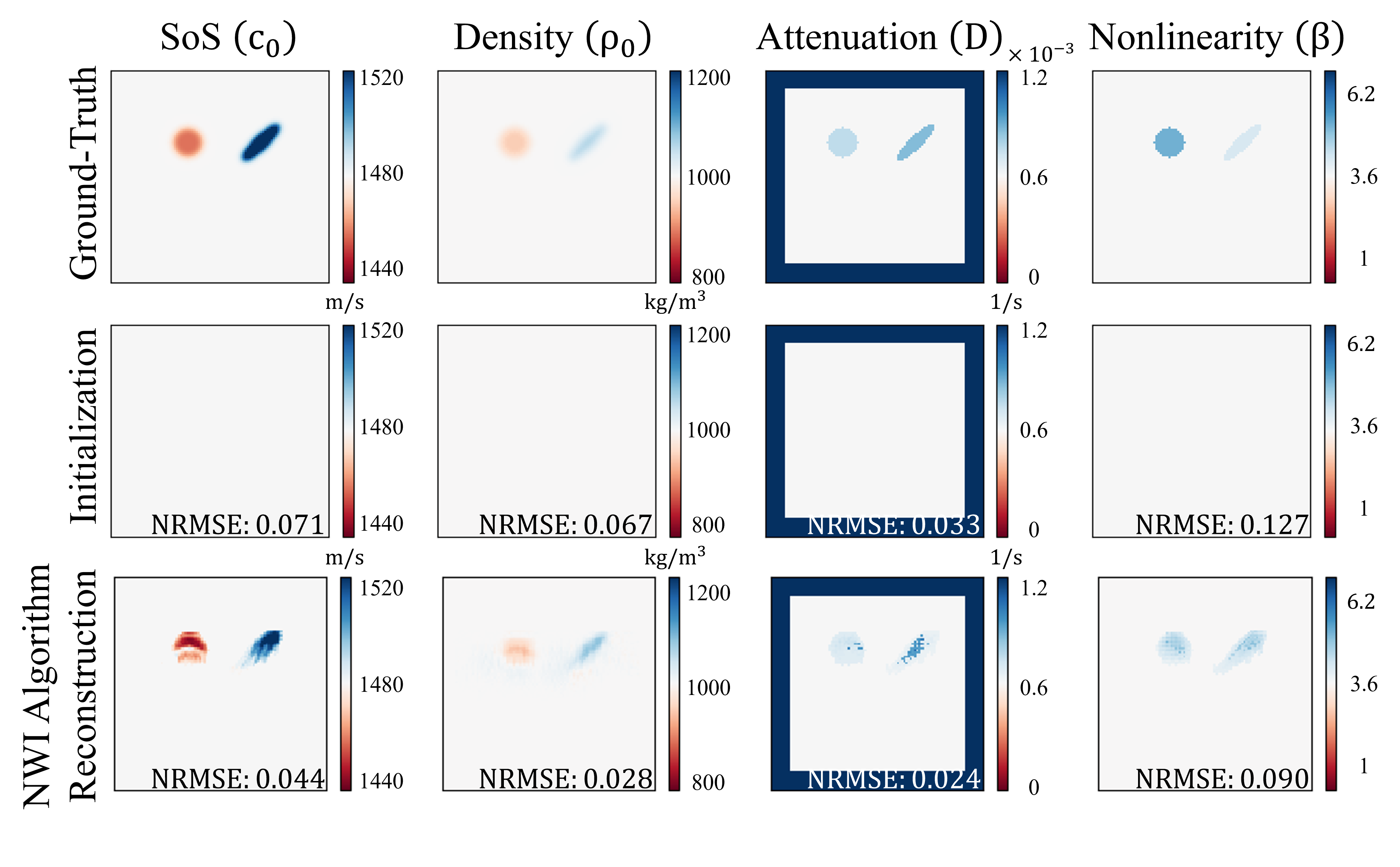}
    \caption{Reconstruction of simulated medium properties, using the NWI algorithm. The first row shows the GT values of the simulated medium. The second row shows the initial values used as input to the inverse algorithm. In the third row we present the obtained reconstructions using NWI. We evaluate the difference using the normalized RMSE metric. Across all maps, white values correspond to the water's properties. The left and right objects are fat and liver respectively.  The PML layer applies a gradual attenuation, which increases close to the boundaries of the grid, as shown in the attenuation maps. }
    \label{fig:Reconstruction}
\end{figure*}

First, we reconstruct the properties of a simulated medium containing two tissues placed in water. The left and right tissues correspond to fat and liver \cite{Duck1990}, as demonstrated in Fig. \ref{fig:Illustrarion}.
This synthetic examples serve serve to asses the correctness of the proposed approach, as was done in previous works \cite{Bernhardt2020,Rau2021}.
In Fig. \ref{fig:Reconstruction}, we show the GT properties of the medium along with their reconstructions. In all maps, the values corresponding to water are marked in white. The PML is the absorbing layer at the grid boundaries with high acoustic attenuation, and is visible in the attenuation maps.

To reconstruct the properties of the simulated medium, we apply Algorithm \ref{alg:algorithm_multiple_pulses} on the sequence of acquired channel data, as detailed in Section \ref{sec:Simulation setup}. The medium is assumed to contain mostly water, and therefore we initialize the properties with values corresponding to water (e.g., the SoS map is initialized to the SoS in water ($1480 \  [m/s]$) and the density map is initialized to water's density ($1000 \ [kg/m^3]$)), as demonstrated in Fig. \ref{fig:Reconstruction}.

Following previous works on SoS estimation \cite{Rau2021,Bernhardt2020}, we use the normalized RMSE evaluation metric to quantitatively evaluate the reconstructions. This metric compares the estimated to the GT properties, and returns values in $[0,1]$:
\begin{equation}
    NRMSE(\hat{\boldsymbol \Theta}_i) = \frac{ \sqrt{\norm{\hat{\boldsymbol \Theta}_i - \boldsymbol \Theta_i^{\star}}^2_F / (n_x  n_z)}}{\boldsymbol \Theta_{i,\max} - \boldsymbol \Theta_{i, \min}},\ \forall i \in [1,n_\theta]
\label{eq:RMSE}
\end{equation}
where $\hat{\boldsymbol \Theta}_i$ are the reconstructed properties, and $\boldsymbol \Theta_{i,\min}$ and $\boldsymbol \Theta_{i,\max}$ are the lower and upper bounds on the property's values, respectively.

In our experiments, the SoS and density were estimated before the damping and nonlinearity parameter. To improve the estimation of the latter, we extract from the estimated density map a mask indicating the tissues' location inside the medium. This is done, by locating the regions in the reconstructed density map, with values that are distant from the density of the water, by a predefined threshold. The estimations of the damping and the nonlinearity parameter are then updated only in a restricted region of the medium, defined by the mask.

\begin{figure*}[t!]
    \centering
    \includegraphics[width=\textwidth]{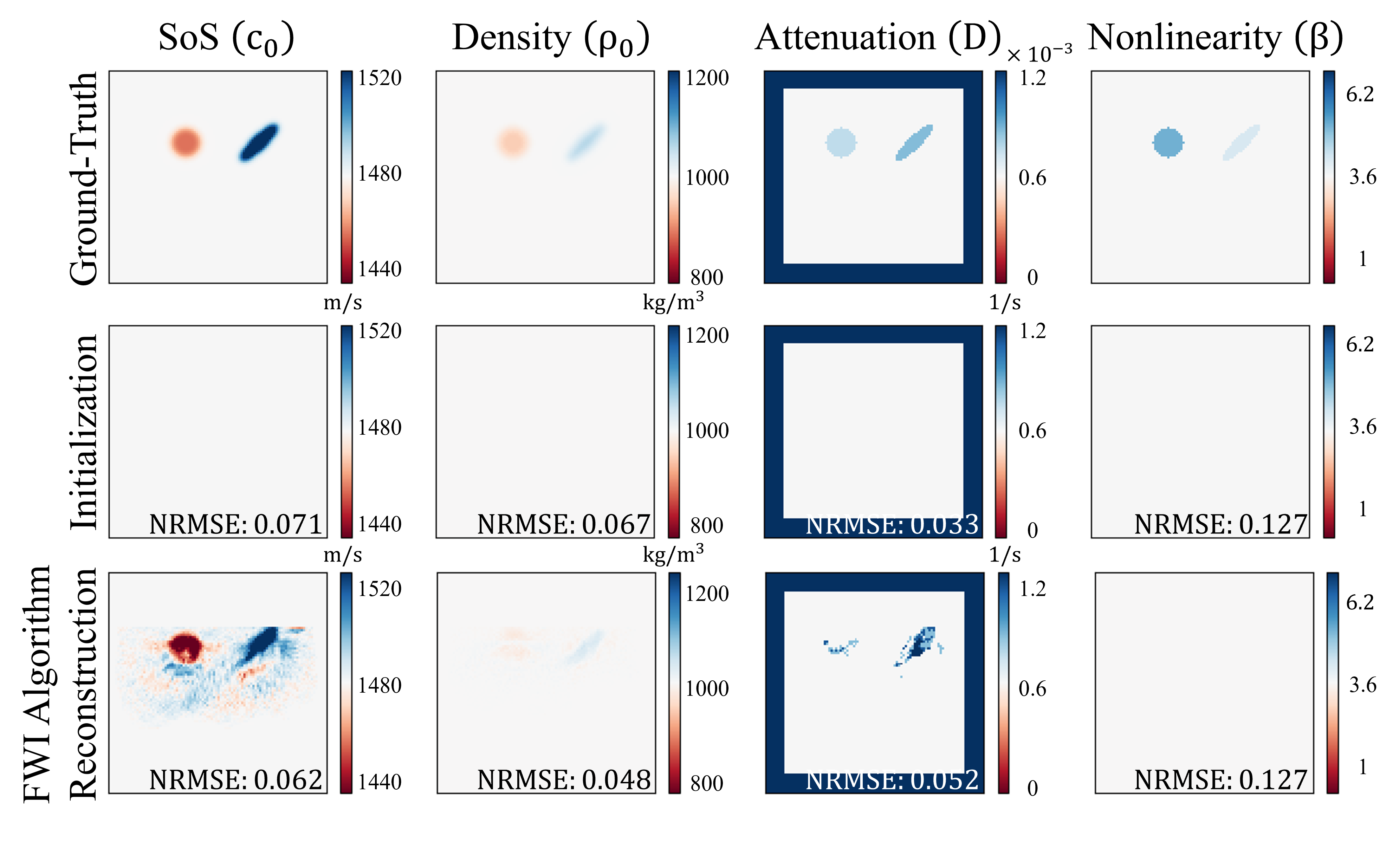}
    \caption{Reconstruction of the SoS, density, and attenuation of the same medium as in Fig. \ref{fig:Reconstruction}, using the FWI algorithm. The first row shows the GT values of the simulated medium. The second row shows the initial values used as input to the inverse algorithm. In the third row we show the results from FWI. The nonlinearity parameter cannot be reconstructed using the linear acoustic model, assumed by the FWI method. The estimation errors of the SoS, density, the attenuation increased by factors of 1.4, 1.7 and 2.2, respectively, highlighting the importance of using an adapted reconstruction algorithm that considers nonlinearity.}
    \label{fig:LA_model}
\end{figure*}

In Fig. \ref{fig:LA_model} we reconstruct the properties of the medium using the FWI algorithm. 
In this case, the nonlinearity parameter, $\mathbf{\beta}$, is not considered in the model, and therefore cannot be reconstructed. The FWI algorithm was tested with the same number of iterations, as the NWI approach, and the same regularization and optimizer were employed during the optimization. We note that the regions consisting of water were estimated incorrectly in the SoS map, which increased its NRMSE score. The estimation errors of the SoS, density, and attenuation increased by factors of 1.4, 1.7 and 2.2, respectively, emphasizing the importance of using an adapted nonlinear acoustic model. Moreover, the reconstructions' time increased substantially.

\section{Discussion}
\label{sec:Discussion}

We presented the NWI algorithm, a model-based approach to reconstruct multiple material's properties in nonlinear media, from US scans. The algorithm exploits the RNN representation of the NLA wave equation to compute the gradients efficiently.

Reconstructing the properties of nonlinear media is suited for medical imaging, where the tissues are highly nonlinear. The nonlinearity poses further challenge compared to linear media, and is not considered in any wave inversion algorithm today. In the example we presented, neglecting nonlinear effects introduced artifacts to the reconstruction, damaged its quality,  and doubled the reconstruction error.

Other imaging modalities are similarly determined by the wave equation, and therefore can be represented as an RNN, such as in photoacoustics and seismology. The proposed method can be applied to these inverse physical problems, by adapting the physical model on which the algorithm relies.

The NWI algorithm expands the range of physical properties that can be reconstructed with US, and improves their reconstruction.  Our results show that considering adapted physical models in nonlinear media, such as the human body, can substantially improve the reconstructions over conventional inverse US methods today. 
Moreover, taking advantage of the RNN representation of the wave equation, reduces time and costs for applying inverse US methods, paving the way for deploying inverse US modalities into the clinic.
In our future work, we hope to conduct an extensive analysis with accurate human tissues simulation, to asses the importance of considering NLA for clinical applications, and explore the advantages of NWI in various clinical applications (including tumors detection and classification, tissues' fat quantification, and imaging of complex media such as cranial US), toward increasing the reliability of US based medical diagnosis.
Also, we aim to generalize the proposed method for reconstructing the properties of three dimensional media, which imposes additional challenges due to the increase in the problem's computational complexity.

\section*{Appendix}

\subsection{Linear Acoustic Operator}
\label{sec:appendix Linear acoustic operator}

In this appendix, we derive the linear acoustic matrix, starting from an homogeneous medium (with constant density). This is a simple model that exemplifies how to construct the linear acoustic operator. Assuming a linear acoustic model in homogeneous media, the wave equation in vector notations reduces to:
\begin{equation}
    \partial_{tt} \mathbf{u} = \text{diag}(\mathbf{c_{0}})^2 (\partial_{xx} + \partial_{zz}) \mathbf{u} + \mathbf{f}
    \label{eq:linear_wave_eq_homogeneous}
\end{equation}
where $\mathbf{c_0} = \text{vec}(\mathbf{C}) \otimes \mathbf{1_v} \in \R^{n_x  n_z n_t \times 1}$ is the discrete medium's SoS organized as a column vector, repeated $n_t$ times, $\mathbf{1_v} \in \R^{n_t \times 1}$ is a vector of ones, and $\text{diag}(\cdot)$ returns a square diagonal matrix with the elements of the input vector on the main diagonal. In addition, $\partial_{tt}$, $\partial_{xx}$, and $\partial_{zz}$ are the discrete second order temporal and spatial derivatives, respectively. This can be written as a linear equation, $\mathbf{A}_h \mathbf{u} = \mathbf{f}$, where
\begin{equation}
    \mathbf{A}_h = \partial_{tt} - \text{diag}(\mathbf{c_{0}})^2 (\partial_{xx} + \partial_{zz}).
    \label{eq:linear_wave_eq_homogeneous2}
\end{equation}
This linear operator is also known as the d'Alembert operator $\mathbf{A}_h = \square$. 
We observe that the linear wave acoustic operator is a function of the medium's SoS, $\mathbf{A}_h = \mathbf{A}_h(\mathbf{c_0})$, and can be represented by an $n_x  n_z  n_t \times n_x  n_z  n_t$ matrix. 
Moreover, the derivative of the operator with respect to the SoS, $\frac{\partial \mathbf{A}_h}{\partial \mathbf{c_0}}$, can be computed, yielding a three dimensional matrix that we do not present here.

We further broaden the linear acoustic operator to capture more complex behaviours, by considering the following linear model, which is similar to (\ref{eq:acoustic_wave_2D}), except neglecting the nonlinearity:
\begin{equation}
\begin{aligned}
    & \partial_{tt} \mathbf{u} + 2 \text{diag}(\mathbf{d})  \partial_{t} \mathbf{u} + \text{diag}(\mathbf{d})^2 \mathbf{u} \\
    & = \text{diag}(\mathbf{c_0})^2 \text{diag}(\mathbf{\rho_0}) \left( \partial_{x} \left(\text{diag}\left(\frac{1}{\mathbf{\rho_0}}\right) \partial_{x} \mathbf{u}\right) \right) \\
    & + \text{diag}(\mathbf{c_0})^2 \text{diag}(\mathbf{\rho_0}) \left( \partial_{z} \left( \text{diag}\left(\frac{1}{\mathbf{\rho_0}}\right) \partial_{z} \mathbf{u} \right) \right) + \mathbf{f}
    \label{eq:acoustic_wave_2D_linear}
\end{aligned}
\end{equation}
where $\mathbf{\rho_0} = \text{vec}(\mathbf{Q}) \otimes \mathbf{1_v},\  \mathbf{d} = \text{vec}(\mathbf{D}) \otimes \mathbf{1_v} \in \R^{n_x  n_z n_t \times 1}$ are the discrete medium's density and attenuation organized as column vectors, respectively, and $\partial_{t}$, $\partial_{x}$, and $\partial_{z}$ are the discrete first order temporal and spatial derivatives, respectively. This can be written as a linear equation, $\mathbf{A} \mathbf{u} = \mathbf{f}$, where 
\begin{equation}
\begin{aligned}
    \mathbf{A} & = \partial_{tt} + 2 \text{diag}(\mathbf{d})  \partial_{t} + \text{diag}(\mathbf{d})^2 \\
    & - \text{diag}(\mathbf{c_0})^2 \text{diag}(\mathbf{\rho_0}) \partial_{x} \text{diag}\left(\frac{1}{\mathbf{\rho_0}}\right) \partial_{x} \\
    & - \text{diag}(\mathbf{c_0})^2 \text{diag}(\mathbf{\rho_0}) \partial_{z} \text{diag}\left(\frac{1}{\mathbf{\rho_0}}\right) \partial_{z}.
    \label{eq:acoustic_wave_2D_linear_operator}
\end{aligned}
\end{equation}
Similarly, this operator depends on the medium's SoS, density, and attenuation, $\mathbf{A} = \mathbf{A}(\mathbf{c_0}, \mathbf{\rho_0}, \mathbf{d})$, and the derivatives, $\frac{\partial \mathbf{A}}{\partial \mathbf{c_0}}, \frac{\partial \mathbf{A}}{\partial \mathbf{\rho_0}}, \frac{\partial \mathbf{A}}{\partial \mathbf{d}}$, can be computed.

Generally, the linear wave equation operator, $\mathbf{A}$, can cover additional behaviours, and depend on additional physical properties, $\mathbf{A} = \mathbf{A}(\left\{ \boldsymbol \theta_i \right\}_{i = 1}^{n_\theta})$. However, we note that this wave representation is limited to linear models, and cannot contain higher orders in $\mathbf{u}$ (such as $\mathbf{u}^2$). 

\subsection{Focused Acoustic Beam}
\label{sec:appendix Acoustic Pulse}

Here we demonstrate how to generate a focused beam from a linear transducer array. We denote by $s(t)$ the acoustic pulse generated from a single transducer element, with a central frequency of $f_0$, and consider a beam created from $N$ transducer elements. We design the beam to focus at a focal point at distance $P$ from the transducer array, on the vertical line from the middle transducer element.

To generate a focused beam at the focal point, all $N$ elements apply the same acoustic pulse, $s(t)$, with tailored time delays. The choice of the acoustic pulse varies in different applications.
The time delays are calculated from the differences between the arrival times of the waves at the focal point, such that the pulses created from all elements, constructively interfere at the focal point \cite{Ganguli2009}. Specifically, the time delay, $\tau_j$, of the $j$th element in the array, $j \in [1,N]$, is:
\begin{equation}
    \tau_j = \frac{P - \sqrt{P^2 + d_j^2}}{\overline{c_0}}
    \label{eq:time delay}
\end{equation}
where $d_j$ denotes the distance between the $j$th element and the middle point in the transducer array. Also, $\overline{c_0}$ is a constant value that characterizes the average SoS inside the medium. Since most of the simulated medium presented in Section \ref{sec:reconstructions} is assumed to be water, we fixed it to the water's SoS $\overline{c_0} = 1480 \ [m/s]$ (in practice, this value is fixed to the average SoS in the human body - $\overline{c_0} = 1540 \ [m/s]$). Accordingly, the acoustic pulse generated from the $j$th element is $s(t - \tau_j)$, for $j \in [1,N]$.
This is sampled with time intervals of $\Delta_t$, and is organized in matrix form such as
\begin{equation}
    \mathbf{F}[x_j, z_j, n] = s(n \Delta_t - \tau_j)
    \label{eq:construct pulse F}
\end{equation}
where $x_j$ and $z_j$ denote the location of the $j$th element on the grid, for $j \in [1,N]$.

\bibliographystyle{IEEEbib}
\bibliography{main}
\end{document}